\documentclass[letterpaper]{article} 
\usepackage[]{aaai25}  
\usepackage{times}  
\usepackage{helvet}  
\usepackage{courier}  
\usepackage[hyphens]{url}  
\usepackage{graphicx} 
\usepackage{natbib}  
\usepackage{caption} 
\usepackage{algorithm}
\usepackage{algorithmic}
\usepackage{amsmath}
\usepackage{amssymb}

\usepackage{newfloat}
\usepackage{listings}
\DeclareCaptionStyle{ruled}{labelfont=normalfont,labelsep=colon,strut=off} 
\floatstyle{ruled}
\newfloat{listing}{tb}{lst}{}
\floatname{listing}{Listing}
\usepackage{multirow}
\usepackage{xcolor}
\nocopyright 

\title{FairVizARD: A Visualization System for Assessing Multi-Party Fairness of Ride-Sharing Matching Algorithms}
\author {
    Ashwin Kumar\textsuperscript{\rm 1},
    Sanket Shah\textsuperscript{\rm 2},
    Meghna Lowalekar\textsuperscript{\rm 3},
    Pradeep Varakantham\textsuperscript{\rm 4},
    Alvitta Ottley\textsuperscript{\rm 1},
    William Yeoh\textsuperscript{\rm 1}
}
\affiliations {
    \textsuperscript{\rm 1} Washington University in St Louis\\
    \textsuperscript{\rm 2} Harvard University\\
    \textsuperscript{\rm 3} IIIT Hyderabad\\
    \textsuperscript{\rm 4} Singapore Management University\\
    ashwinkumar@wustl.edu, sanketshah@g.harvard.edu, meghna.lowalekar@iiit.ac.in, pradeepv@smu.edu.sg, alvitta@wustl.edu, wyeoh@wustl.edu
}

\begin{document}

\maketitle

\begin{abstract}
There is growing interest in algorithms that match passengers with drivers in ride-sharing problems and their fairness for the different parties involved (passengers, drivers, and ride-sharing companies). Researchers have proposed various fairness metrics for matching algorithms, 
but it is often unclear how one should balance the various parties' fairness, given that they are often in conflict. We present FairVizARD, a visualization-based system that aids users in evaluating the fairness of ride-sharing matching algorithms. FairVizARD presents the algorithms' results by visualizing relevant spatio-temporal information using animation and aggregated information in charts.  FairVizARD also employs efficient techniques for visualizing a large amount of information in a user friendly manner, which makes it suitable for real-world settings. We conduct our experiments on a real-world large-scale taxi dataset and, through user studies and an expert interview, we show how users can use FairVizARD not only to evaluate the fairness of matching algorithms but also to expand on their notions of fairness.
\end{abstract}

\maketitle
\sloppy


\section{Introduction}\label{introduction}




On-demand taxi ride-sharing systems (e.g.,~UberPool and Lyft Line) have seen a surge in popularity in many countries worldwide.
Unlike conventional (non-shared) on-demand taxi systems, where a taxi driver is driving a single passenger (or a single group of passengers) from one location to another, in a ride-sharing system, a taxi driver is driving multiple passengers, each with different pickup and dropoff locations. 
To improve the adoption of such systems, \emph{Ride Hailing Companies} (RHCs) include incentives for all \emph{parties} in the system. Passengers are incentivized by having a discount on the fare of their ride compared to a traditional non-shared ride. Drivers are incentivized by having an increase in earnings due to the larger number of passengers 
they can drive in each trip. And RHCs are incentivized by having an increase in the number of passengers requesting rides, leading to an increase in the revenue of the system.

\begin{figure*}[t]
\centering
    \includegraphics[width=0.4\linewidth]{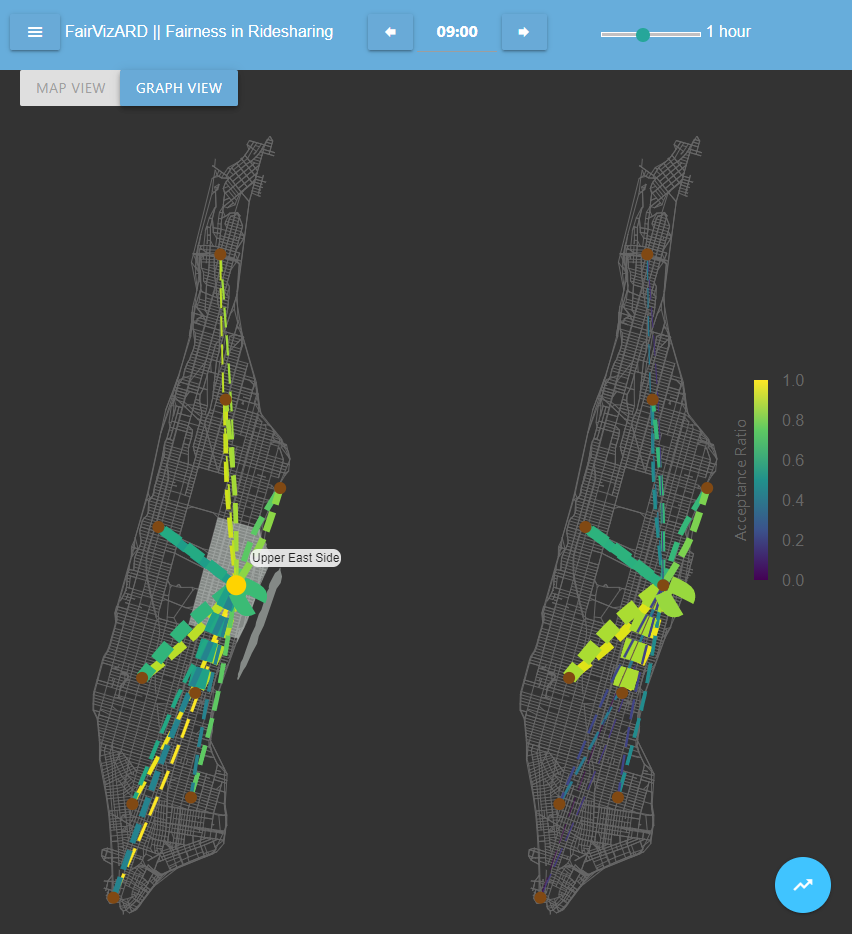}
    \includegraphics[width=0.4\linewidth]{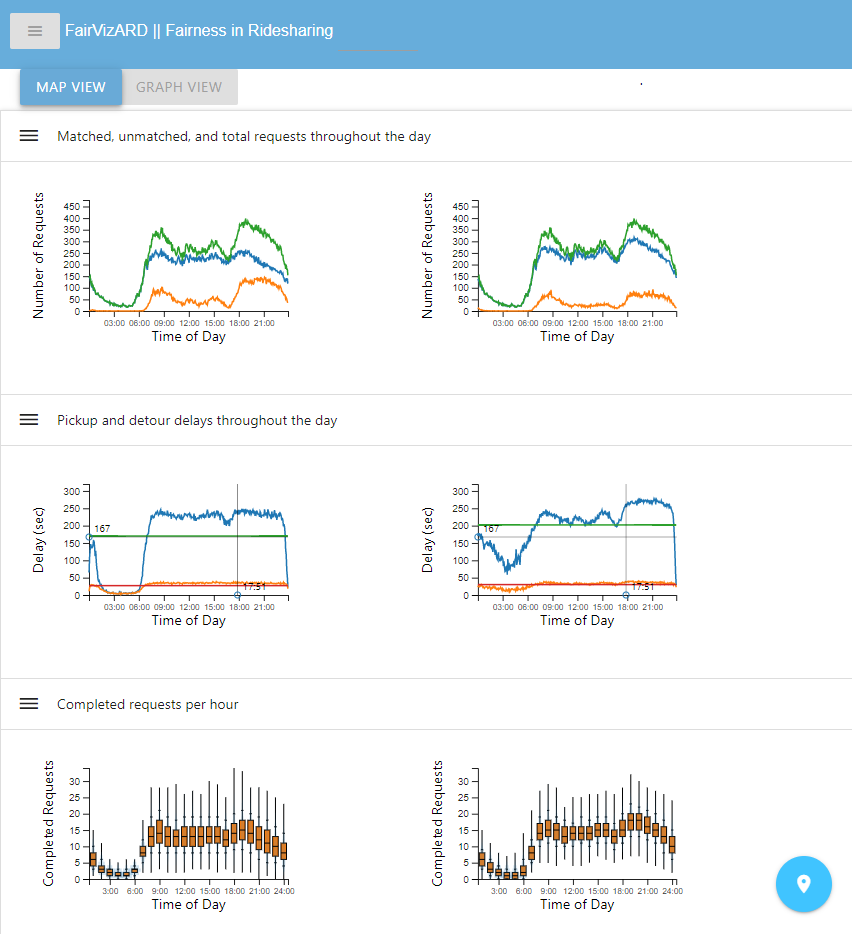}
    \caption{From left to right: (i)~\emph{Map View} of FairVizARD. (ii)~\emph{Graph View} of FairVizARD. 
    } 
    \label{fig:teaser}
\end{figure*}

However, the possibility of a taxi driver getting requests from multiple passengers at different pickup and dropoff locations introduces an additional layer of complexity to \emph{matching algorithms} that RHCs employ to match requests by passengers to taxi drivers. For example, when considering matching a request to a driver with passengers in their taxi, the matching algorithm will now need to account for the necessary change in the route of the taxi and, consequently, the additional delay added to the trip of passengers in the taxi.
Researchers have proposed a number of matching algorithms for ride-sharing systems by optimizing different objectives (e.g., minimizing the length of the detour of passengers, maximizing the average number of requests served by drivers, etc.) while keeping the runtimes of the algorithms small to ensure computational efficiency~\cite{ride_alonso,ride_huang2014,ride_ma2015}. 

A key limitation with existing matching algorithms is that they mostly optimize for monetary factors 
or user satisfaction. 
As a result, the matches proposed can be \emph{unfair} (i.e,~they have underlying biases that disadvantage certain parties of the system). For example, a matching algorithm that keep drivers in busy areas to maximize their revenue while ignoring requests from passengers in other areas is biased in favor of drivers and against passengers in non-busy areas. 
Consequently, there is growing interest in analyzing and improving the fairness of matching algorithms~\cite{rf_passenger_cost_benefit_sharing_2016,rf_Balancing_eff_fairness_lesmana2019,rf_tradeoff_nanda2020}.
Existing work often distills the notion of fairness into a single quantifiable metric that matching algorithms can optimize. While this is a reasonable goal, we are curious if there is a consensus on an agreed upon notion of fairness and how one should trade off fairness between the different parties in a ride-sharing system. 
This is especially important given that there are multiple parties using such a system and are impacted by matching algorithms that RHCs employ.

Towards this end, we present \textbf{FairVizARD}, a \textbf{Fair}ness \textbf{Viz}ualization for \textbf{A}nalysis of \textbf{R}ide-sharing \textbf{D}ata (see Figure~\ref{fig:teaser}).\footnote{FairVizARD was presented as a System Demonstration at ICAPS 2021.} 
Using a combination of map- and graph-based visualizations, FairVizARD shows spatio-temporal information of the passenger requests and taxi drivers as well as statistics on different factors that may be of interest to different parties in the system. Ridesharing systems deal with a large number of taxis and passenger requests. So, it is also important for the visualization system to be able to present a large amount of spatio-temporal information in a compact yet user-friendly way. FairVizARD uses efficient techniques to present all the information required by users to study the fairness of matching algorithms for different parties while reducing the information density on the map. To allow users to easily compare different matching algorithms and understand their tradeoffs, FairVizARD presents this information for two algorithms side by side, where both algorithms are given the same initial problem configuration. 
To the best of our knowledge, FairVizARD is the first visualization system that enable users (as well as algorithm designers) to analyze and compare city-scale ride-sharing matching algorithms for fairness and other performance metrics. 


\section{Background and Related Work}\label{sec:related-work}


\noindent \textbf{Ride-sharing Matching Problems:}
While there are many variants of \emph{ride-sharing matching problems} (RMPs),
in a typical problem definition, a matching algorithm receives as inputs a continuous stream of batches of requests from passengers $\mathcal{U}$ containing their pickup and dropoff locations and the current locations of all the taxis $\mathcal{D}$. It then needs to match as many requests to taxis as possible in a way that optimizes some objective function  (e.g.,~maximizes the number of requests matched) subject to physical constraints $\kappa$ (e.g.,~maximum number of passengers that can share a taxi) and time-related constraints $\mathcal{T}$ (e.g.,~passengers with matched requests must be picked up within a certain time).

More formally, a typical RMP is defined by a tuple $\langle \mathcal{G}, \mathcal{C}, \mathcal{D},\mathcal{U}, \kappa, \mathcal{T}, \Delta \rangle$, where:
\begin{itemize}
\item $\mathcal{G} = \langle L, E \rangle$ is a graph representing a road network, where each node $l \in L$ is a location in the city and each edge $e \in E$ connects two neighboring nodes in the graph.
\item $\mathcal{C}: E \rightarrow \mathbb{R}^+$ is a function that defines the cost $\mathcal{C}(e)$ of each edge $e \in E$ in the graph. 
\item $\mathcal{D}$ is a set of taxi drivers in the problem. 
\item $\mathcal{U}$ is the set of requests from passengers, with each request $u_i \in \mathcal{U}$ defined by the tuple $\langle s_i,g_i,t_i,f_i\rangle$ containing the pickup location $s_i \in L$, the dropoff location $g_i \in L$, the arrival time $t_i$ of the request, and the fare $f_i$ that will be paid if the request is served.
\item $\kappa$ is the maximum number of passengers that can share one taxi. 
\item $\mathcal{T}$ is the set of time-related constraints imposed on the problem. Examples include the \emph{pickup delay}, defined as the difference between the arrival time of a request and the time the passenger is picked up, or the \emph{detour delay}, defined as the extra time taken to reach a dropoff location due to ride-sharing, being within some user-defined upper bound. Both of these delays are computed using the edge costs $\mathcal{C}(e)$ of the relevant edges taken by the taxi. 
\end{itemize}

A \emph{solution} $M = \{(u_i,d_j), \ldots\}$ is a set of matches between request $u_i \in \mathcal{U}$ and taxi driver $d_j \in \mathcal{D}$ such that all the constraints in $\kappa$ and $\mathcal{T}$ are satisfied. A solution is \emph{optimal} if it optimizes its objective function, which could differ across problem definitions. 

RMPs has been extensively studied, where researchers have introduced methods that improve the quality of the matches made in terms of increasing the number of requests matched~\cite{ride_zac,ride_zacbenders,ride_ma2015}, reducing the pickup and detour delays~\cite{ride_alonso, ride_huang2014, ride_ma2015}, and increasing the revenue of the drivers~\cite{rf_Balancing_eff_fairness_lesmana2019}. The complexity of RMP algorithms and, as a result, the time taken by algorithms to match drivers to passenger requests, increases with the increase in the value of $\kappa$. As the runtime of the RMP algorithms need to be relatively small, most existing works have either considered assigning one request at a time (sequentially) to available drivers for high value of $\kappa$ ~\cite{ride_ma2015,ride_tong2018unified,ride_huang2014} or assigning all active requests together in a batch for a small value of $\kappa$~\cite{ride_yu2019integrated,ride_zheng2018order}. The sequential solution is faster to compute but the solution quality is typically poor~\cite{uberblog}. Researchers have proposed integer optimization approaches for assigning all active requests together for a high value of $\kappa$~\cite{ride_alonso}, and further improved them by including information about anticipated future requests while matching the current batch of requests~\cite{ride_neurADP,ride_zacbenders}. 

\smallskip \noindent \textbf{Fairness in Ride-sharing:}
Early work in this area has focused on fairness from the perspective of passengers, where researchers assume that each passenger will get a discount on the fare of their ride depending on the type of \emph{partners} that share the taxi. With this assumption, researchers investigated how the discounts should be distributed across passengers~\cite{rf_passenger_cost_benefit_sharing_2016}.
Later work more formally defined the concept of ride-sharing fairness while tackling the issue of transparency by allowing passengers to indicate their preferences for the type of partners they wish to share their rides with~\cite{rf_FairVOpt_wolfson2017}. 
When such preferences are available, researchers can then apply game-theoretic concepts such a stable pairings~\cite{rf_nash_foti2019}. 
More recent work has focused on fairness from the perspective of drivers as well as two-sided fairness for both passengers and drivers~\cite{rf_2side_fair_suhr2019}. 
There is also work on group-based fairness that aims at identifying and balancing bias against certain groups of passengers or drivers on grounds like location, race, and gender~\cite{rf_tradeoff_nanda2020,rf_tradeoff2_xu2020}.

The literature thus far has focused only on defining and quantifying fairness in such a way that they can be used as objective functions or additional constraints of RMP algorithms. 
In contrast, we are interested in developing a visualization system that can be used to better understand the implicit notions of fairness of laypersons using ride-sharing systems. None of the works discussed thus far 
have visualized the fairness of RMP algorithms.

\section{FairVizARD Design}\label{system}

We now discuss the design of FairVizARD, including the design challenges, the various components of the visualization system and the new fairness metric we propose, \emph{Zonal Fairness}.
Our goal was to build an interactive visual interface that allows users to explore different notions of fairness and, as a bonus, also allows algorithm developers to analyze the performance of their algorithms across common metrics. To that end, we had the following design challenges:
\begin{itemize}
    \item \textbf{Identifying key points of information to visualize:}
    With so many possible factors affecting notions of fairness, one of the key challenges was to identify and prioritize the information to visualize. Since the data is primarily spatio-temporal data, we decided to use a map-based visualization (\emph{map view}) to describe the spatial distribution of the data and a graph-based visualization (\emph{graph view}) to describe the temporal distribution of the data.
    
    \item \textbf{Handling a large volume of data:}
    Another key challenge was finding out suitable aggregation techniques to handle the large volume of data points that needed to be visualized. It was more of a challenge on the map view, with over over 200,000 requests coming in over a day, and 1,000 drivers active at any time based on the algorithms used to simulate ride-sharing. One of the primary ways we tackled this issue was by using binning. Data is aggregated spatially (e.g., within zip codes or neighborhoods) and temporally before being presented to the user.
    In addition, when multiple layers of visualization are active at the same time, we reduce the information density on the screen by suppressing the visibility of some of the features until users interact with them.
    
    \item \textbf{Accounting for multiple parties or stakeholders of the system:}
    Since the goal of FairVizARD is to allow the exploration of fairness across multiple stakeholders (i.e.,~passengers, drivers, and RHCs), it was important to visualize data relevant to all stakeholders. Based on metrics found in the literature for ride-sharing fairness, we identified important pieces of information to display for the different stakeholders, which we describe in the following subsections. 
    
\end{itemize}

FairVizARD shows data for a single day (24 hours) with one-minute resolutions. This resolution is decided based on the decision epochs considered by the algorithms we run, and can be scaled as required. We demonstrate FairVizARD using Manhattan, NY as the network graph. All algorithms were run using demand data from the NY Yellow Taxi Dataset~\cite{yellowtaxi}, with 1000 drivers operating across 24 hours. FairVizARD can be used to look at a single algorithm at a time, or to compare two or more algorithms side by side in a split-screen view.

\subsection{Map View}\label{map view}


Recall that the data to visualize is primary spatio-temporal data. Therefore, FairVizARD is composed of two views: A \emph{map view} that uses a map-based visualization to visualize the spatial distribution of the data and a \emph{graph view} that describe the temporal distribution of the data. We now describe the map view and discuss the graph view later.

Figure~\ref{fig:teaser}(i) shows the map view displaying a specific type of information, which we will discuss later. In general, this view shows data at a particular time of day distributed across the map. 
When aggregated data is shown, then users can also adjust the \emph{time window} over which the data is being aggregated by using a slider (Figure \ref{fig:teaser}(i)). 
Users can select different kinds of information to display on this view through the use of a sidebar. Each of these sub-views can be stacked, allowing users to look for correlations across different aggregation methods and data. Figure~\ref{fig:map-view-examples} illustrates the four possible sub-views:

\begin{figure}[t]
  \centering
   \includegraphics[width = \linewidth]{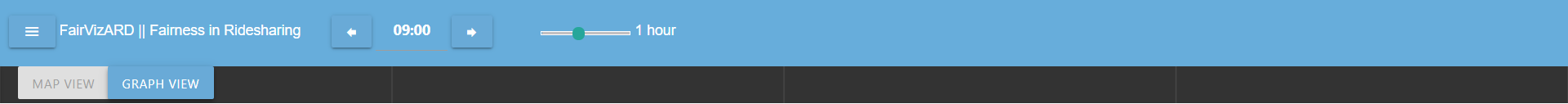}
  \includegraphics[width = 0.22\linewidth]{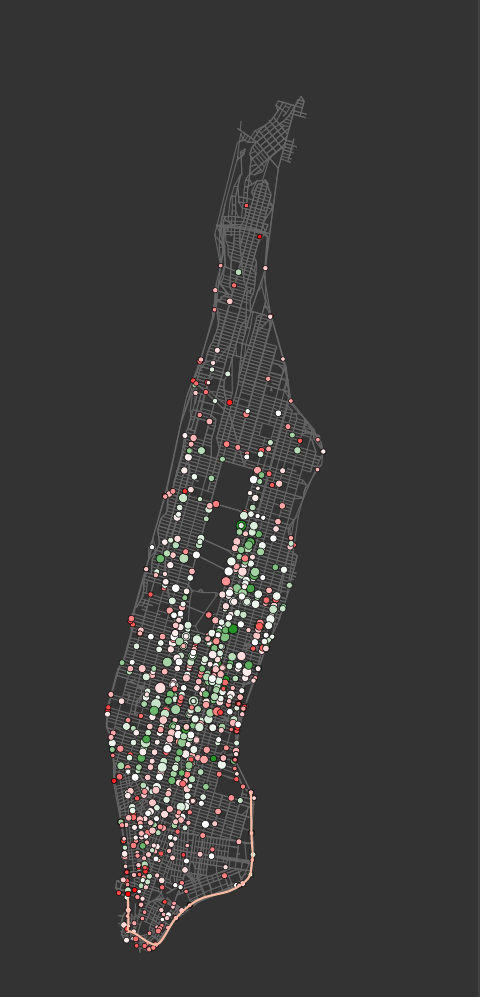}
  \includegraphics[width= 0.22\linewidth]{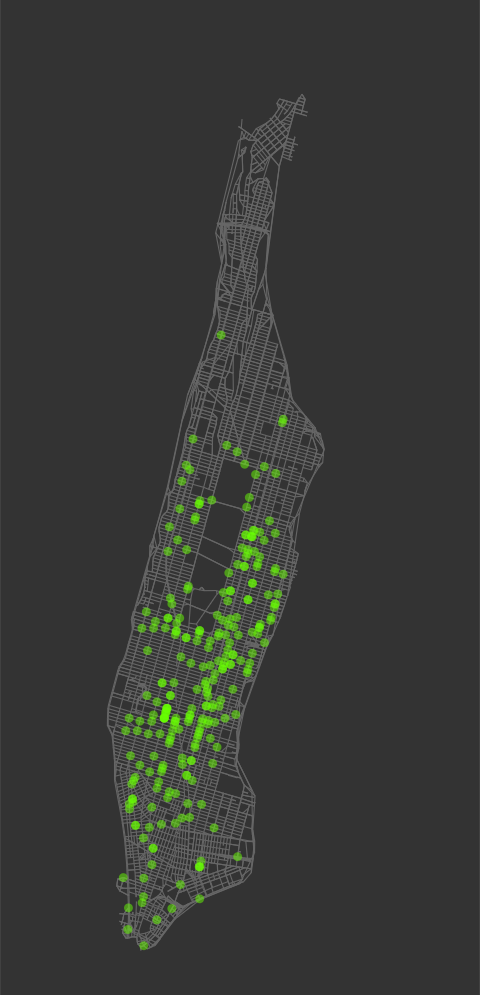}
  \includegraphics[width= 0.22\linewidth]{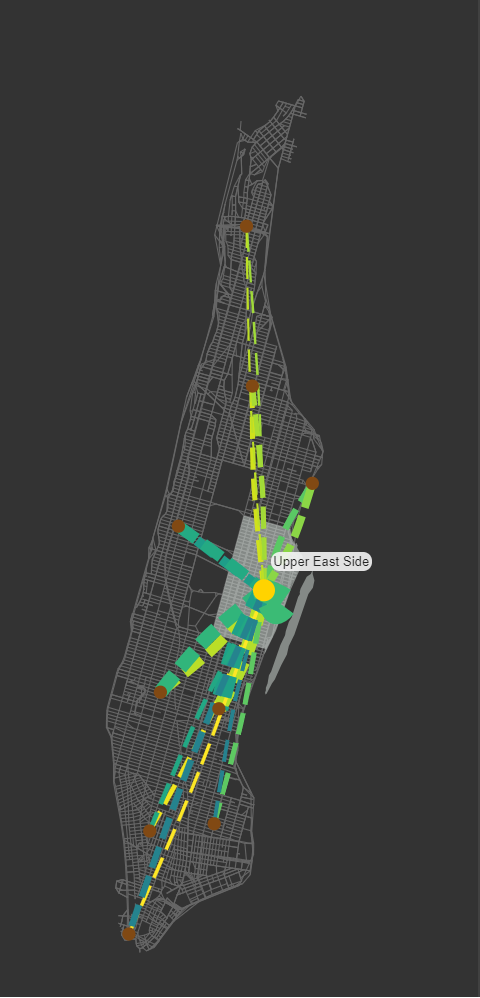}
  \includegraphics[width= 0.22\linewidth]{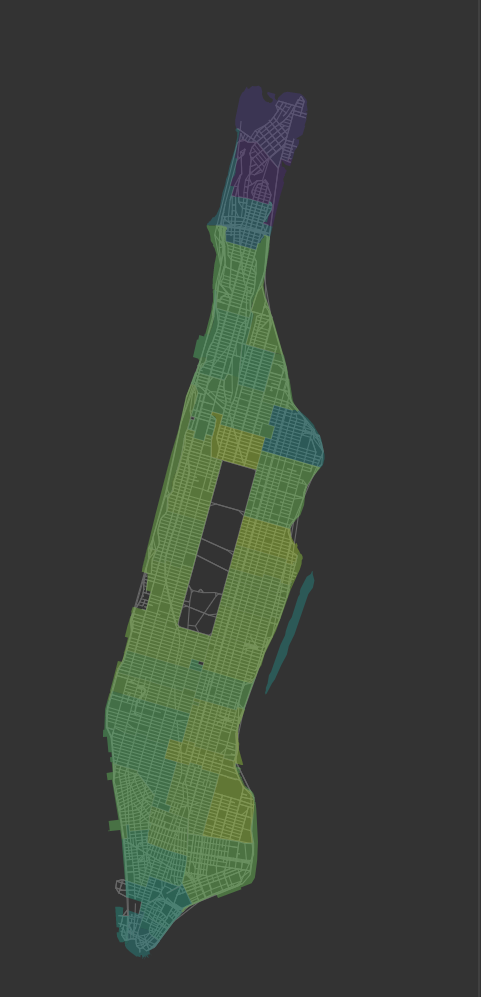}
  \includegraphics[width = 0.066\linewidth]{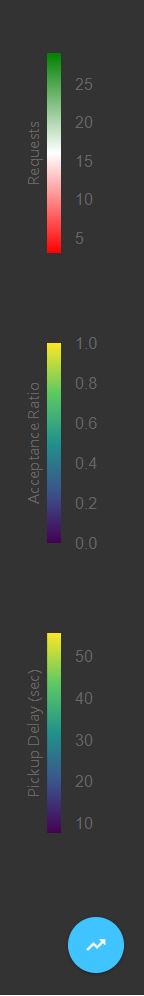}
  \caption{Possible Map Views. From left to right: (i) Taxi Data; (ii) Request Data; (iii) Inter-Zone Data; (iv) Individual Zone Data; (v) Legend}  
  \label{fig:map-view-examples}
\end{figure}

\begin{itemize}
\item \textbf{Taxi Data:} Figure~\ref{fig:map-view-examples}(i) shows the locations of all taxis at the current time, where each individual circle corresponds to a taxi. The size of a circle grows proportionally to the number of active requests of the taxi at the current time step. The color of a circle indicates the number of requests matched to the taxi over the previous time window, where the legend is shown on the top of Figure~\ref{fig:map-view-examples}(v). We use a diverging color scale of red-white-green because it may be of interest to spot outliers that are matched with too few requests (colored red) or too many requests (colored green) compared to the average.
Clicking on a circle highlights the path of the taxi (colored orange), showing the pickup and dropoff locations of its active requests along the way. This sub-view was intended to give users a quick overview of the distribution of taxis over the city, and to help understand the spread of requests over drivers.

\item \textbf{Request Data:} Figure~\ref{fig:map-view-examples}(ii) shows the pickup locations of all the requests received at the current time step as green circles. Users can also toggle on the dropoff locations of those requests, which will be shown as red circles. Additionally, users can also filter the data to just show the matched requests, unmatched requests, or all requests. This sub-view thus give users a sense of where passengers are currently located, where they want to travel to, and whether requests from some location or requests to some location are consistently unmatched by the algorithm.

\item \textbf{Inter-Zone Data:} Figure~\ref{fig:map-view-examples}(iii) shows the relationship between zones in the city, where zones are contiguous regions that are either defined by zip codes or by user-defined neighborhoods. 
Each zone is visualized on the map as a node located at the centroid of the zone. A directional edge (shown by a flow animation) from zone $z_1$ to zone $z_2$ indicates the number of matched requests, whose pickup locations are in zone $z_1$ and dropoff locations are in zone $z_2$, in the previous time window. The thickness of the edge is proportional to the square root of the number of matched requests. 
The color of an edge from zones $z_i$ to $z_j$ can be used to indicate one of two possible metrics:
\begin{itemize}
\item [$\circ$] \emph{Acceptance ratio}, which is the ratio between the number of matched requests $|\mathcal{M}(z_i, z_j)|$ from zone $z_i$ to zone $z_j$ and the total number of incoming requests $|\mathcal{U}(z_i,z_j)|$ from zone $z_i$ to zone $z_j$. Figure~\ref{fig:map-view-examples}(iii) colors the edges using this metric, where the legend is shown in the middle of Figure~\ref{fig:map-view-examples}(v). 

\item [$\circ$] \emph{Average detour delay}, which is the average detour delays for all matched requests from zones $z_i$ to $z_j$. 
\end{itemize}

Users can hover over a zone and the visualization will only show the incoming and outgoing edges of that zone and the geographical boundary of that zone. For example, Figure~\ref{fig:map-view-examples}(iii) shows the visualization when the user is hovering over the ``Upper East Side'' zone. This sub-view was intended to allow users to analyze the amount of traffic between zones and the degree in which requests from/to a zone is not matched or have disproportionately long delays. 

\item \textbf{Individual Zone Data:} Figure~\ref{fig:map-view-examples}(iv) illustrates data related to zones, which are either defined by zip codes or user-defined neighborhoods. Each zone is colored according to its \emph{average pickup delay}, which is the average difference between the arrival time of matched requests from the zone and the time the passenger is picked up. 
This sub-view allows users to identify neighborhoods with disproportionately long delays. 
\end{itemize}

Through the design of the map view, we also discover a new fairness metric, called \emph{Zonal Fairness}, that is fairly intuitive. Using the same notion of zones as described above, the zonal fairness value $Z_f$ is defined as:
\begin{align}
    Z_f  =  \min_{(z_i, z_j)} AR(z_i,z_j) = \min_{(z_i, z_j)} \frac{|\mathcal{M}(z_i,z_j)|}{|\mathcal{U}(z_i,z_j)|}
\end{align}
where $(z_i,z_j)$ is an arbitrary pair of zone nodes, and $AR(z_i,z_j)$, $\mathcal{M}(z_i,z_j)$, and $\mathcal{U}(z_i,z_j)$ are the acceptance ratio, number of matched requests, and number of incoming requests from zone $z_i$ to zone $z_j$, respectively. 
Intuitively, the value of $Z_f$ indicates the ratio of the requests matched for the pair of zone nodes that is most underserved. The value of $Z_f$ can be extracted from the Inter-Zone Data sub-view if the user chooses to visualize the acceptance ratio metric. A hypothetical matching algorithm could also use this metric as its objective function and seek to maximize $Z_f$.

\begin{figure}[t]
  \centering
  \includegraphics[width=0.49\linewidth]{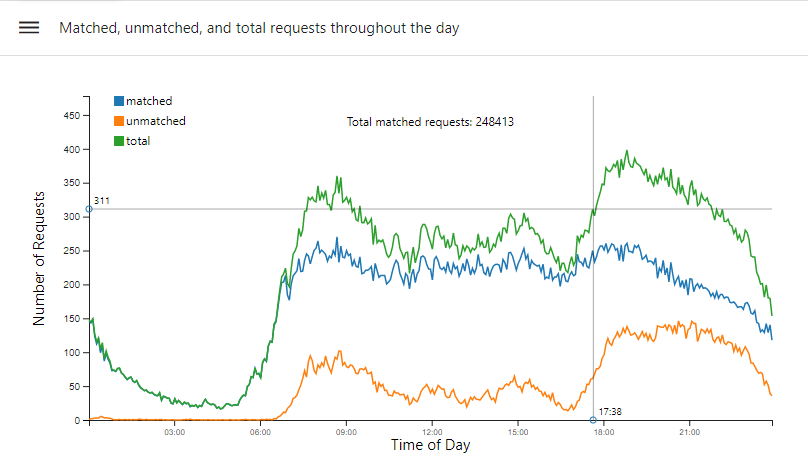}
  \includegraphics[width=0.49\linewidth]{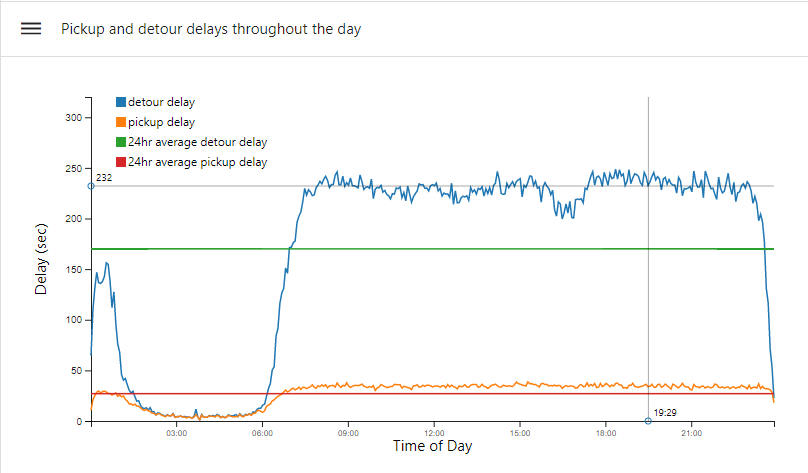}
  \includegraphics[width=0.49\linewidth]{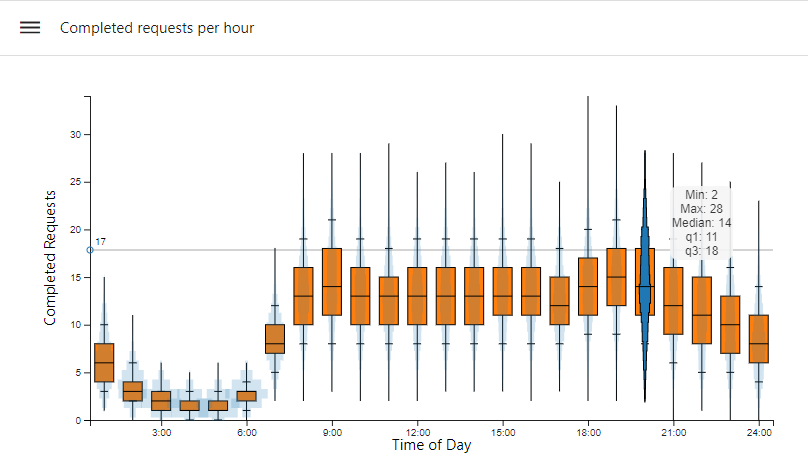}
  \includegraphics[width=0.49\linewidth]{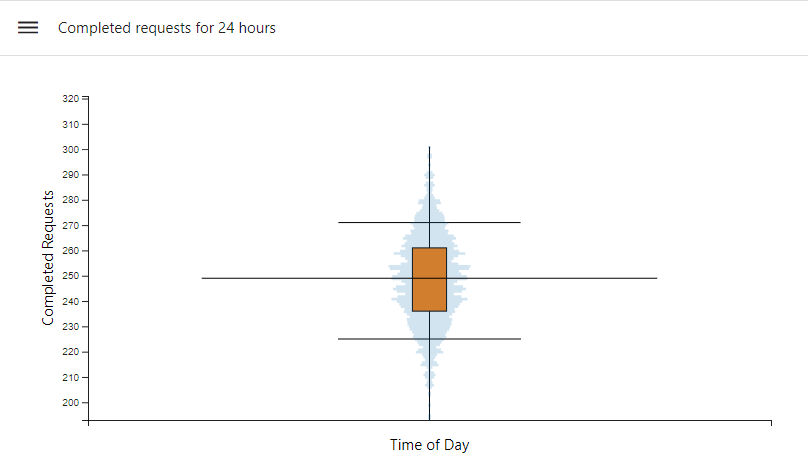}
  \caption{Examples of various plots in the graph view. Clockwise, from top left: (i) Incoming Request Graphs; (ii) Pickup and Detour Delay Graphs; (iii) Completed Request Graphs (per hour); (iv) Completed Request Graphs (per day)}  
  \label{fig:graph-view-examples}
\end{figure}


\subsection{Graph View}\label{graph view}




We now discuss the graph view of FairVizARD, which visualizes temporal distribution of the data. 
Figure~\ref{fig:graph-view-examples} displays the four possible graphs that are shown in graph view of FairVizARD. Each graph is under its own collapsible header.
The types of graphs are as follows:
\begin{itemize}
\item \textbf{Incoming Request Graphs:} The graphs at the top of Figure~\ref{fig:teaser}(ii) show examples, where they plot the number of matched, unmatched, and total requests received against the time of the day that the requests were received. Users can use these graphs to identify peak times in the day and to get a sense for when requests are usually unmatched.

\item \textbf{Pickup and Detour Delay Graphs:} The graphs in the middle of Figure~\ref{fig:teaser}(ii) show examples, where they plot the average pickup and detour delays for requests arriving at each time step as well as their 24-hour averages. Users can use these graphs to evaluate algorithms based on metrics that are important to passengers. Using these graphs in conjunction with request graphs will also allow users to identify correlations between the number of requests received and the delays, if they exist.

\item \textbf{Completed Request Graphs:} The graphs at the bottom of Figure~\ref{fig:teaser}(ii) show examples, where they plot box plots of the distribution of the number of requests completed by taxi drivers throughout the day. 
Each box plot provides the range of requests completed (i.e.,~their minimum and maximum values), the median (shown by the horizontal line in the orange rectangle), the second and third quartiles (the ranges corresponding to the bottom and top portions of the rectangle, respectively), and the 10th and 90th percentiles (shown using ticks on the vertical line). 
Users can use these graphs to evaluate algorithms based on metrics that are important to drivers. For example, if the box plots show a large variance during some time of the day, then it means that some drivers are disproportionately completing fewer requests compared to others.
\end{itemize}

\section{Evaluation Setup}\label{sec:setup}

We now discuss the setup for our evaluation, where we compared FairVizARD against a benchmark visualization of numeric metrics through exploratory user studies and an expert interview. The goal of the evaluation is to examine how FairVizARD impacts data comprehension and exploration of ride-sharing matching algorithms. We had the following goals in our user study:
\begin{itemize}
    \item \textbf{Goal 1 (User Understanding of Data):} Understand if FairVizARD allows users understand the same data as effectively as a set of contextually-relevant numeric metrics used in literature to quantify fairness.
    \item \textbf{Goal 2 (Facilitation of Exploration):} Discern if users could make fairness-related observations through FairVizARD that they could not make through numeric metrics alone. 
    \item \textbf{Goal 3 (Accommodation of Different Notions of Fairness):} Better understand the diversity of opinions on fairness and see if FairVizARD allows users to explore their notions of fairness.
\end{itemize}

\begin{figure*}[t]
    \centering
    \includegraphics[width = 0.32\linewidth]{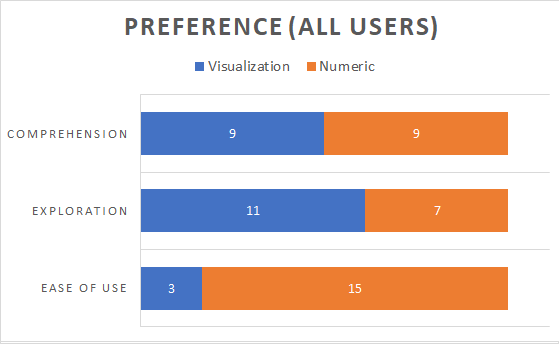}
    \includegraphics[width = 0.32\linewidth]{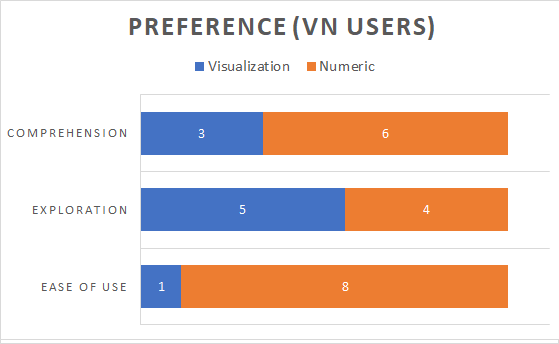}
    \includegraphics[width = 0.32\linewidth]{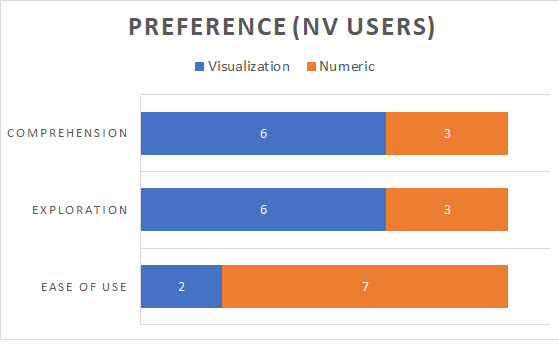}
    \caption{System preference across users, on the basis of ease of use, exploration, and comprehension of fairness. From left to right: (i)~Preferences of all users; (ii)~Preference of users in the VN ordering; (iii)~Preference of users in the NV ordering
    }
    \label{fig:System preference across users}
\end{figure*}

\subsection{User Study Design}


We ran an exploratory user study with 18 participants, where each participant used both FairVizARD and a dashboard of numeric metric values. 
Participants completed a survey where they needed to rank two ride-sharing matching algorithms in terms of their fairness for each of the three parties (i.e.,~passenger, driver, and RHC) in the system. In the survey, the participants were also asked to describe their own notions of fairness that they were using to rank the algorithms and describe how they used the information shown in each system to measure the two algorithms based on their notions of fairness.


\smallskip \noindent \textbf{Numeric Metrics as Benchmark Comparison:}
The current state of the art in evaluating the fairness of matching algorithms is through numeric values of different fairness metrics. For example, researchers have seeked to maximize the minimum revenue of taxi drivers~\cite{rf_Balancing_eff_fairness_lesmana2019} and to improve the equitability in the distribution of revenue of taxi drivers~\cite{rf_2side_fair_suhr2019}.
We thus developed a dashboard that presents relevant statistical information of common fairness metrics for the algorithms being evaluated. Specifically, it presents information for the following metrics: (1)~Statistical information on the distribution of completed requests across all taxi drivers; (2)~Total number of completed requests; (3)~Statistical information on the distribution of acceptance ratio across all minutes of a day; (4)~Statistical information on the distribution of inter-zone acceptance ratio across all pairs of zones; (5)~Statistical information on the pickup delay across all matched requests; and (6)~Statistical information on the detour delay across all matched requests.

\smallskip \noindent \textbf{Accounting for Ordering Bias and Learning Effects:} Since all users will use both FairVizARD and the dashboard of numeric metrics to compare the matching algorithms, the ordering in which they use the systems may influence their perceptions of the systems. For example, users may have more difficulty using the first system compared to the second system since they are already more familiar with the data by the time they use the second system. Therefore, we randomly split our pool of participants into halves, where one half uses FairVizARD before the dashboard and vice versa for the other half, to better understand and account for ordering biases. 
We hypothesize that the ranking of the two algorithms for a user should remain unchanged across both FairVizARD and the dashboard of numeric metrics. However, to ensure that users did not carry over their observations and conclusions from using the first system to the second system, we swapped the positions of the algorithms and used different input data for the two systems.

\subsection{Participants}

We recruited 18 volunteers (13 men and 5 women), whose ages are between 23 and 45 years, with an average age of 28 years. All participants had some graduate education and had taken an algorithms course at some point in time. 17 participants reported familiarity with ride-sharing systems. We use the identifiers VN1 to VN9 to refer to the 9 participants that use FairVizARD first before using the dashboard of numeric metrics, and we use identifiers NV1 to NV9 to refer to other 9 participants who used the dashboard first.\footnote{We used `N' for \emph{numeric metrics} and `V' for \emph{visualization}.}

\subsection{Ride-sharing Algorithms}

We evaluated FairVizARD using two ride-sharing matching algorithms (RewardPlusDelay and NeurADP),
whose objective is to maximize the number of requests matched and both enforce the same capacity and time-related constraints. 
\emph{RewardPlusDelay}~\cite{ride_alonso} is a myopic algorithm that does not consider the future value of matches while making matches in the current time step. After generating the feasible groups of requests, it matches taxis to these groups using an integer linear program. Aside from the objective of maximizing the number of requests matched, it also adds a component to the objective function to prefer matches that minimize detour delays.

\emph{Neural Approximate Dynamic Programming} (NeurADP)~\cite{ride_neurADP} is a non-myopic approach that consider the future value of matches while making matches in the current time step. It does so by using a neural network to learn the expected future value (i.e.,~the expected future effect of current matches of taxis to feasible groups of requests) and incorporates those learned expected future values in its integer linear program. Aside from the objectives of RewardPlusDelay, it also seeks to maximize the expected future value in its objective function.


\section{Evaluation Results}
\label{sec:results}

\begin{table*}[t]
\caption{Measures used by users to define their notions of fairness; some users provided more than one measure}
{\small
\begin{center}
\begin{tabular}{|l|l|c|}
\hline
\textbf{Party} & \textbf{Fairness Measures Defined by Users} & \textbf{\# Users} \\
\hline
\hline
  \multirow{9}{*}{Drivers}& Equal revenue  & 7\\
 
 & High average revenue &  7\\
 & All drivers in same area/time should get the same number/type of matched requests & 3\\
 & Certainty that they will get matched throughout the day/Consistency in getting matches & 2\\
 & Travel time to pick up passenger should be small & 2\\
 & Minimize idle hours & 1\\
 & Guaranteed revenue (i.e., a base salary) & 1\\
& Maximize distribution of taxis across different locations & 1\\
& Getting matched based on their preference (e.g., preferred destinations) & 1\\
\hline 
\hline 
 \multirow{7}{*}{Passengers} & Minimize average detour delay & 14\\
 & Uniform acceptance ratio across locations & 11\\
 & High acceptance rates/Low rejection rates for all locations & 4\\
 & Uniform detour delay across locations & 3\\
 & Requests from the same pickup location and time should have the same likelihood of getting matched & 1\\
 & Identical detour delays for requests with the same pickup and dropoff locations & 1\\
 & Minimize maximum detour delay & 1\\
\hline 
\hline
\multirow{5}{*}{Ride-}& Maximize profit/requests matched & 15\\
\multirow{5}{*}{Hailing} & Balance between driver-side fairness and passenger-side fairness & 9 \\
\multirow{5}{*}{Companies}& Smaller detour delay & 2\\
& Maximize number of shared rides & 2\\
& Matching requests across different locations & 1\\
& Maximize passenger satisfaction & 1\\
& Reputation & 1\\
\hline
\end{tabular}
\end{center}
}
\label{table:fairness_measures}
\end{table*} 

We now discuss the results of our evaluations, first, in the context of the three goals of our study.

\smallskip \noindent \textbf{Goal 1 (User Understanding of Data):} To evaluate the degree to which a user understands the data presented to quantify fairness, we count how many of the following four key easily-observable differences between the two algorithms they observed: (1)~NeurADP matched more requests compared to RewardPlusDelay, which corresponds to more revenue for RHCs. (2)~The variance in the number of matched requests per taxi driver is smaller with NeurADP than with RewardPlusDelay. (3)~RewardPlusDelay matched more requests from zones with smaller number of requests. (4)~RewardPlusDelay's matched requests have smaller average delays than NeurADP's matched requests.

Overall, users using FairVizARD and the dashboard with numeric metrics observed on average 3.38 and 3.22 differences, respectively. There was no ordering bias observed since users using FairVizARD observed at least as many differences as when they use the dashboard in both orderings. Therefore, we conclude that FairVizARD does allow users to understand data that is relevant for fairness as easily as, if not better than, numeric metrics.

\smallskip \noindent \textbf{Goal 2 (Facilitation of Exploration):} In general, users were able to make fairness-related observations using FairVizARD that they did not make using the dashboard with numeric metrics. We provide an anecdotal observation below of user NV1, who used the dashboard first before using FairVizARD. When using the dashboard, they noted that ``Neighbourhoods are more equally served" for RewardPlusDelay. When using FairVizARD, they noticed underserved neighborhoods with low acceptance ratios for NeurADP. Overall, they concluded that RewardPlusDelay is fairer:
\begin{quote}
    ``\emph{To me fairness would mean serving all neighborhoods and all people equally. It is more important than maximizing income for drivers or companies. Algorithm 2 underserves Inwood and Harlem, where there are more people of color and black people. For this reason I think Algorithm 1 is better.}''
\end{quote}
where Algorithms~1 and~2 correspond to RewardPlusDelay and NeurADP, respectively. The fact that the user knew the population demographics in the different areas of Manhattan, coupled with having the data spatially represented on a map, allowed them to make more nuanced fairness-related observations than if they used numeric metrics alone.

\smallskip \noindent \textbf{Goal 3 (Accommodation of Different Notions of Fairness):} The diversity of opinions on fairness among our users is apparent, as shown by Table~\ref{table:fairness_measures}, where most users defined multiple notions of fairness for a single party. Given the myriad notions of fairness, even for a single application domain of ride-sharing, it becomes even clearer that it is unlikely that a single fairness metric can capture all the underlying intricacies involved in measuring the fairness of a matching algorithm. 
Therefore, there is a strong need for exploratory systems that allow users to interact with data across different dimensions to come to a more well-informed conclusion.

\subsection{Ordering Effects}
\label{sec:ordering-effects}
\begin{figure}[t]
\centering
    \includegraphics[width=0.8\columnwidth]{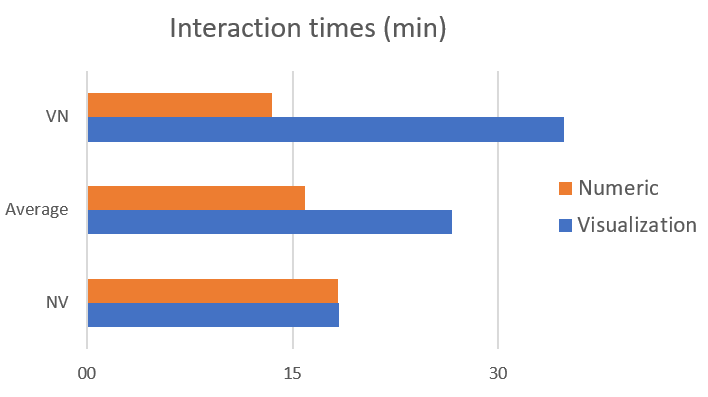}
    \caption{Average user interaction times}
    \label{fig:time_spent}
\end{figure}

We also observed that the order in which users used the systems influenced their preferences and time spent on each system (see Figures~\ref{fig:System preference across users} and~\ref{fig:time_spent}). 

\smallskip \noindent \textbf{Ordering Effect on Preference:} Overall, both FairVizARD and the dashboard with numeric metrics are identical in terms of their \emph{ease of comprehension}. However, users preferred whichever system they used second, probably because they were already more familiar with the data. 
Unsurprisingly, users also thought that it is \emph{easier to explore} the data using FairVizARD, though the ratio of users who thought so is higher in the group that used the dashboard with numeric metrics first compared to the group that used FairVizARD first. Again, we suspect that users were not able to as easily explore the data because they were overwhelmed with FairVizARD as their first system. In contrast, when users are already familiar with the broad trends in the data after using the dashboard, they are able to better investigate finer trends through FairVizARD. 
Finally, also unsurprisingly, users also strongly preferred the dashboard of numeric metrics in terms of the \emph{ease of use}. This preference is visible in both groups, likely because the dashboard requires significantly less effort to find relevant data, almost giving users direct answers to the questions.

\smallskip \noindent \textbf{Ordering Effect on Time Spent:} On both orderings, users spent more time on FairVizARD than on the dashboard with numeric metrics (see Figure~\ref{fig:time_spent}). However, the difference is marginal when users used the dashboard first and is substantially larger when they use FairVizARD first. As described above, once users are familiar with the data after using the dashboard, they are able to more easily use FairVizARD. The NV group of users finished the study 9 minutes faster than users in the VN group, which is a significant difference.

\subsection{Qualitative Comments}
We now discuss qualitative comments provided by users, where we group them thematically and provided representative samples from each group:

\smallskip \noindent \textbf{Quick vs. In-Depth Analysis:} The biggest difference between FairVizARD and the dashboard with numeric metrics is the volume of information presented to the user. Given that the dashboard provide more concise aggregate information, many users found it easier to use:
\begin{quote}
    ``\emph{Numeric fairness metrics provide less information at once but make it easier for viewer to have a idea with less time.}'' -- VN6
\end{quote}
Two users commented that the curation of the metrics meant that statistics that were hard to see in FairVizARD were easier to make sense of with the dashboard; they both used FairVizARD first before using the dashboard. On the flip side, users believe that FairVizARD allows for a more in-depth analysis through better exploration of the data and the creation of a `more informed decision':
\begin{quote}
    ``\emph{The visualization revealed more information. It took some time to understand all the functions but once I understood, I felt like I was able to investigate more and better.}'' -- NV1
\end{quote}

The consensus is that numeric metrics concisely convey `relevant' information for a quick analysis while FairVizARD allows for in-depth analyses. This suggests that the difference in preference may fall roughly along the lines of how users prefer to interact with their data -- summary vs in-depth. This, in turn, could be a reflection of the `need for cognition' of different users \cite{cacioppo1996dispositional}. This position is nicely summarised by the following quote:
\begin{quote}
 ``\emph{Visualization provided far more information and nuance, but at the expense of having to critically think about what it means to benefit a stakeholder.}'' -- VN5
\end{quote}

\smallskip \noindent \textbf{Kinds of Information:}
Another theme in the comments was about the utility of different types of information in the two systems. An important feature of numeric metrics was the presence of curated aggregate information. The absence of this in FairVizARD is a con for many users:
\begin{quote}
    ``\emph{There was not a clear way to see the total zone-wise acceptance rate of rides because the visualization required me to put in a specific time. If there was a visualization that summarized that information for the whole day, that would have been a lot more useful.}'' -- VN2
\end{quote}
On the other hand, the presence of spatio-temporal information was important to many users:
\begin{quote}
    ``\emph{The visualizations highlighted the potential disparate treatment among the different neighborhoods, which I didn't think of during Task A.}'' -- NV4
\end{quote}
As a result, we believe that a combination of both systems is the way forward. We posit, however, that while the FairVizARD could easily be modified to include the aggregate information, it would be difficult to include that volume of spatio-temporal information numerically.

\subsection{Expert Interview}

In addition to the user study, we also conducted a semi-structured interview with an expert who has worked in the ride-sharing domain for 12 years in areas like fleet optimization, driver behavior analysis, agent-based simulation of taxis, labor economics (of drivers), and impact of ride-hailing innovation. We thus leveraged their experience to get better insights on the utility of FairVizARD, by asking them some analytical questions. Their responses to the various questions are discussed below.

They were able to identify all four key easily-observable differences discussed in Goal 1 above. 
Similar to a majority of users, they also felt that the dashboard of numeric metrics is easier to use:
\begin{quote}
    ``\emph{For the purpose of reaching high-level judgement (e.g., fairness), numeric fairness metrics allows well-trained decision makers to quickly arrive at conclusions}.''
\end{quote}
Both for ease of comprehension and ease of exploration, they ranked the visualization as the better system, where FairVizARD allows people to:
\begin{quote}
``\emph{Actually explore and spot microscopic trends (particularly spatial-temporal patterns on the map); the true value of the visualization is [that] you don't need to know beforehand what exactly you're looking for}.''
\end{quote}
An improvement for FairVizARD can be to include the numeric metrics, since they are useful in different scenarios: 
\begin{quote}
    ``\emph{For a high level decision maker, numeric metrics are what they are more used to particularly if they are transportation expert. Visualization is good, but they probably only need those numbers, the key performance statistics, then they can make decisions}.''
\end{quote}

\begin{quote}
    ``\emph{Let's say there is a major change, in demand pattern, in driver population, or the fleet size, then I think the visualization will help them to think about, for example, new metrics they should come up with to better manage or better regulate}.''
\end{quote}

\subsection{Utility of FairVizARD for Algorithm Developers}\label{discussion_alg}
During the course of development of the system, we also observed that FairVizARD can help ride-sharing algorithm developers in identifying and resolving issues in their systems.  

Ride-sharing systems are complex large scale spatio-temporal systems. There are thousands of taxis in the system and hundreds of customers arrive at each decision epoch. The matching decision taken by algorithms at one decision epoch has an impact on the decisions at future decision epochs. Due to such complications, the systems are hard to debug and silent errors are common. The visualization can provide aid in debugging such systems by showing the locations of taxis and customers and their movement across time. 

While analyzing fairness of NeurADP algorithm using FairVizARD, we observed that after reaching certain intersections, some taxis were not moving from their location. We communicated this observation to NeurADP developers and they found that the issue occurred due to a bug in their simulator where if an empty taxi reaches an intersection from where the time taken to move to any neighboring intersection is more than decision epoch duration, it was not moving. The visualization helped NeurADP developers uncover this bug in their simulator and it improved the performance of their algorithm by nearly 5\%. 
\subsection{Discussions}\label{discussion}

Overall, we found that FairVizARD performed well on all the three goals of the user study. The results show that it allowed users to (Goal~1)~make fairness-related observations from the data; (Goal~2)~not only better explore the data, but also integrate their background knowledge, to arrive at more nuanced fairness-related conclusions; and (Goal~3) have their diverse notions of fairness accommodated. 
An analysis of the fairness metrics that users used highlighted the diversity in opinions on fairness in ride-sharing (see Table~\ref{table:fairness_measures}). Additionally, users focused more on average-case statistics instead of the worst case, which is important because the literature on fair algorithms often seek to improve the worst case. 
Finally, when users are balancing the fairness of different parties, they tend to try to be fair to the majority.

A natural question that arises then is the following: `If FairVizARD met all its goals, why did the user preferences not reflect it?' We believe that the answer is two-fold: (1)~Users have an inherent preference on the amount of effort necessary to reach conclusions from the data. The qualitative comments suggest that the difference in preference between the visualization and numeric metrics is more a reflection of this than about how each system allowed users to process fairness. (2)~The current iteration of FairVizARD isn't perfect. Although the visualization provides spatio-temporal information, it does not provide aggregate information that frames the problem using statistics initially shown to the user (in the case of NV participants) and it does not allow for macro-level comparisons between different algorithms.
From these observations and expert user recommendations, we conclude that an ideal system would involve a combination of both high-level statistics from the dashboard of numeric metrics and in-depth information provided by FairVizARD. Further, from the ordering effects observed, presenting the high-level statistics first is the way to go.

This leads us to our final point of discussion -- the use of FairVizARD by algorithm developers. Though we did not highlight this in the paper, the visualization is a powerful tool that can also be used to find silent errors related to the matching algorithm analyzed. 
Our expert user went a step further and suggested that algorithm developers and even high-level decision makers can use such a system during anomalous events when metrics calibrated for the day-to-day may no longer be as relevant.

\section{Conclusions}

As ride-sharing becomes increasingly pervasive and algorithms are used to match passenger requests to taxi drivers, it is important to understand the impact of those algorithms in terms of the fairness for the different parties in the system. Existing work has focused on distilling the notion of fairness into a single quantifiable metric that can be optimized. However, users have diverse opinions on fairness, as illustrated by Table~\ref{table:fairness_measures}. 
Therefore, in this paper, we proposed FairVizARD with the goal of allowing users to better analyze matching algorithms based on their notions of fairness. Our user studies confirmed that users were able to better explore the data presented when using FairVizARD compared to using a dashboard with numeric metrics; both FairVizARD and the dashboard are equally easily comprehensible; and the dashboard is easier to use compared to FairVizARD. In general, users agreed that FairVizARD provided more information and at a higher level of detail, but at the expense of having to think critically about how to make sense of that data in the context of fairness. In conclusion, FairVizARD allowed users to investigate other notions of fairness that may not be possible with existing metrics through more thorough explorations of the data.

In the future, we plan to expand FairVizARD’s capabilities to support additional tasks such as pricing comparisons, visualization of cancellation patterns, and a unified interface that enables direct interaction with the visualization from within the simulation environment. These enhancements aim to provide users with an even richer toolkit for exploring the multifaceted nature of fairness in ride-sharing systems. We also intend to incorporate visualizations for recent advances in fair ride-sharing algorithms released after this study~\cite{rf_neurADP_fair2020, SI_kumar2023}, thereby ensuring that FairVizARD remains a relevant and adaptable platform for both researchers and practitioners interested in fairness-aware decision making.


\bibliography{refs}
\end{document}